\documentstyle[aps,prl,floats,psfig]{revtex}
\catcode`@=11
\def\references{%
\ifpreprintsty
\bigskip\bigskip
\hbox to\hsize{\hss\large \refname\hss}%
\else
\vskip 24pt
\hrule width\hsize\relax
\fi
\list{\@biblabel{\arabic{enumiv}}}%
{\labelwidth\WidestRefLabelThusFar  \labelsep4pt %
\leftmargin\labelwidth %
\advance\leftmargin\labelsep %
\ifdim\baselinestretch pt>1 pt %
\parsep  4pt\relax %
\else %
\parsep  0pt\relax %
\fi
\itemsep\parsep %
\usecounter{enumiv}%
\let\p@enumiv\@empty
\def\theenumiv{\arabic{enumiv}}%
}%
\let\newblock\relax %
\sloppy\clubpenalty4000\widowpenalty4000
\sfcode`\.=1000\relax
\ifpreprintsty\else\small\fi
}
\catcode`@=12
\draft

\def\mh{m_h^{}}
\def\gev{\rm GeV}
\def\fbi{\rm fb^{-1}}
\def\ww{WW^*}
\def\mt{\mu\tau}
\def\kmt{\kappa_{\mu\tau}}
\def\lsim{\mathrel{\raise.3ex\hbox{$<$\kern-.75em\lower1ex\hbox{$\sim$}}}}
\def\gsim{\mathrel{\raise.3ex\hbox{$>$\kern-.75em\lower1ex\hbox{$\sim$}}}}

\def\ll{\ell\bar\nu \bar\ell \nu}

\begin{document}

\twocolumn[\hsize\textwidth\columnwidth\hsize\csname
@twocolumnfalse\endcsname

\hfill\vbox{
\hbox{MADPH--00--1188}
\hbox{hep-ph/0008141}}

\title {$h\to \mu\tau$ at Hadron Colliders}
\author{Tao Han and Danny Marfatia}
\address{Department of Physics, University of Wisconsin--Madison, WI 53706}

\maketitle

\begin{abstract} 
We study the observability for a lepton flavor-changing decay 
of a Higgs boson $h\to \mu\tau$ at hadron colliders. 
 Flavor-changing couplings of a Higgs boson exist at tree
level in models with multiple Higgs doublets. The $h\mu\tau$
coupling is particularly motivated by the favorable interpretation 
of \mbox{$\nu_\mu-\nu_\tau$} oscillation.
We find that at the Tevatron Run II
the unique $\mu\tau$ signature could serve as the Higgs 
discovery channel, surpassing expectations for Higgs boson
searches in the SM and in a large parameter region of the MSSM. 
The sensitivity will be 
greatly improved at the LHC, beyond the coverage at a muon 
collider Higgs factory. 
\end{abstract}
]

The standard model (SM) of electroweak interactions
and many of its extensions generically predict the existence
of Higgs bosons. Detecting Higgs bosons and studying 
their properties in future collider experiments 
would provide crucial information for the mechanism of 
electroweak symmetry breaking and hopefully 
fermion flavor physics as well. These have been the
most prominent issues in contemporary particle physics.

The upgraded Fermilab Tevatron will start its mission
next year with c.~m.~energy $\sqrt s=2$ TeV 
and an annual luminosity $L\approx 2~\fbi$ 
per detector (Run IIa). Ultimately,
one would hope to reach an integrated luminosity
of $L\approx 15-30~\fbi$ (Run IIb). 
In terms of the search for the SM Higgs boson ($h$), 
the most promising processes beyond the LEP2 reach would be electroweak 
gauge boson-Higgs associated production \cite{scott}
$
p\bar p \to W h,\ Zh.
$
The leptonic decays of $W,Z$ provide a good trigger and 
$h\to b\bar b$ may be reconstructible with adequate $b$-tagging
and $b\bar b$ mass resolution, allowing a Higgs boson reach
of $m_h\sim 120-130$ GeV \cite{runii}.
For a heavier Higgs boson $m_h\approx 2M_W$, the leading
production channel via gluon fusion $gg\to h$ 
and the relatively clean decay mode
$h\to WW^* \to \ell\bar\nu \bar\ell \nu$ may be useful
in digging out a weak Higgs boson signal \cite{tevprl}.
It is believed that a SM-like Higgs boson may be observable 
up to a mass of about 180 GeV at a $3\sigma$ statistical 
level for $L\approx 25~\fbi$ \cite{runii}. In the minimal  
supersymmetric extension of the standard model (MSSM),
the mass of the lightest CP-even Higgs boson is bounded
by $m_h\lsim 130$ GeV \cite{mhbound}. When the CP-odd 
Higgs state ($A$) of the MSSM is heavy $m_A \gsim 2M_Z$, the
lightest Higgs boson has SM-like properties and the conclusion
for a light SM Higgs boson search remains valid 
in a large parameter region of the MSSM.  
The only exception is when $m_A\sim {\cal O}(M_Z)$
and $\tan\beta$ (ratio of the Higgs
vacuum expectation values) is large, where the 
production of $b\bar b h,\ b\bar b A$ is enhanced 
by $\tan^2\beta$ and $h,A\to b\bar b, \tau\bar \tau$
may be accessible \cite{mssmh}.
At the CERN Large Hadron Collider (LHC) with $\sqrt s=14$ TeV
and $L\approx 100-300~\fbi$, one expects to fully cover the
range of theoretical interest \mbox{$m_h\lsim 1$ TeV} for the SM
Higgs boson, or to discover at least one of the MSSM Higgs 
bosons \cite{LHC}.

The Higgs sector is the least constrained in theories
beyond the SM. It is thus prudent to keep an open mind when
studying Higgs physics phenomenologically and experimentally. 
A particularly important question about the Higgs sector 
is its role in fermion flavor dynamics, {\it i.e.}, 
the generation of fermion masses and flavor mixings. There
have been attempts to explain flavor mixings by
a generalized Higgs sector with multiple Higgs doublets.
It is argued \cite{model3} that the fermion flavor mixing structure
due to the Higgs coupling at tree level can be of the form,
\begin{equation}
\kappa_{ij}\ {\sqrt{m_i m_j} \over v}\ h^0 {\bar \psi_i} \psi_j\,,
\label{coupling}
\end{equation}
where $i,j$ are generation indices and $v\approx 246$ 
GeV is the Higgs
vacuum expectation value. $\kappa_{ij}$ is a product of 
the model parameter  $\lambda_{ij}$ and the neutral Higgs 
mixing $\cos\alpha$ \cite{model3}. Although they are free parameters
without a priori knowledge of a more fundamental
theory, $\lambda_{ij}$ is naturally order of unity
from a model-building point of view and $\cos\alpha=1$ 
corresponds to no Higgs mixing. Such Higgs-fermion
couplings would yield flavor-changing neutral currents,
and therefore lead to rich phenomenology \cite{pheno,g2,hdecay,sher}. 
However, transitions involving the light generations are
naturally suppressed and the largest couplings occur
between the third and second generations. 

In this Letter we explore the lepton flavor-changing
coupling  $\kappa_{\mu\tau}$ of a Higgs boson. This is 
particularly motivated by the favorable interpretation 
for $\nu_\mu-\nu_\tau$ flavor oscillation from recent 
atmospheric neutrino experiments \cite{superK}. 
If a large mixing between $\nu_\mu$ and $\nu_\tau$ exists
as indicated by the Super-K experiment \cite{superK},
then it will necessarily lead to the decay $h \to \mu \tau$.
The branching fraction depends on the particular
model of the Higgs sector, which can be parameterized
by $\kappa_{ij}$. The current
constraints on this coupling from low energy experiments
are rather weak, giving $\lambda_{\mu\tau}<10$ 
derived from the muon anomalous magnetic moment \cite{g2}.
Other low energy probes are not expected to be sensitive enough 
to reach the natural size $\lambda_{\mu\tau}\sim {\cal O}(1)$.
The potentially interesting lepton flavor-changing decay modes 
for a Higgs boson were recently discussed \cite{hdecay}, 
and their search at a muon collider \cite{hfactory} 
was studied \cite{sher}. In this work,
we propose to look for the signal at the upgraded Tevatron
and the LHC. The leading production mechanism for a neutral
Higgs boson through gluon fusion is 
\begin{equation}
p p(\bar p)\to ggX\to h X\to \mu\tau X.
\label{process}
\end{equation}
We find that due to the unique flavor-changing signature
and the distinctive kinematics of the signal final state, 
the Tevatron Run II will have significant sensitivity to
such a coupling, making this signal a possible Higgs
discovery channel for $m_h \approx 100 - 140$ GeV
if $\kmt\sim {\cal O}(1)$.
At the LHC, the sensitivity is substantially
improved leading to a probe for the coupling to
a level of $\kappa_{\mu\tau}\sim 0.15$ and extending
the mass coverage to 160 GeV.

\vskip 0.1in
\noindent
\underline{$h$ production and decay at hadron colliders}

\begin{figure}[tb]
\mbox{\psfig{file=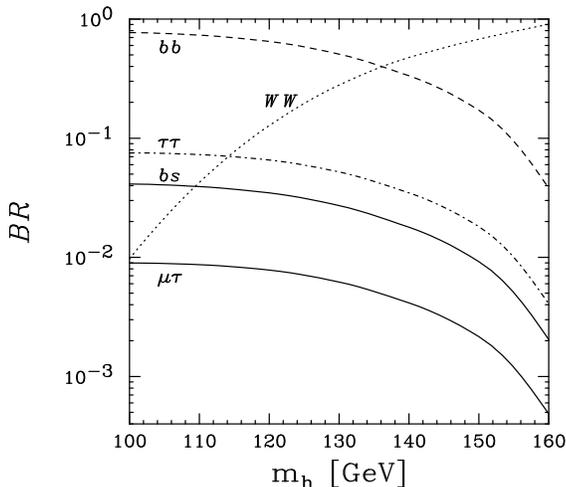,angle=90,width=7.5cm,height=6.5cm}}
\caption{The Higgs boson decay branching fraction versus $m_h$.  
The coupling parameters $\kappa_{ij}$ are taken to be one.}
\label{BR}
\end{figure}

The dominant decay mode for a SM-like Higgs boson is
$h\to b\bar b$ for $m_h<130$ GeV and $h\to WW^*$ 
for a heavier mass.
The partial decay width for $h\to \mu\tau$ is given by
\begin{equation}
\Gamma(h\to \mu\tau)={\kappa_{\mu\tau}^2\over 4\pi}
{m_\mu m_\tau\over v^2}\ m_h\,.
\end{equation}
Here and henceforth $\mu\tau\equiv\mu^-\tau^+ + \mu^+\tau^-$.
In comparison to the $\tau^+\tau^-$ mode in the SM, we have 
\mbox{$\Gamma(h\to \mu\tau)/\Gamma(h\to \tau\tau)=2\kmt^2(m_\mu/m_\tau)$.}
In Fig.~\ref{BR}, we show these decay branching fractions
versus the Higgs boson mass. The $\mt$ mode is plotted 
assuming $\kmt=1$, for which BR($h\to \mu\tau$) is at the 
$1\%$ level. For $\kmt \approx 3$, the $\mt$ mode can
be as large as the SM $\tau^+\tau^-$ mode. For $m_h>140$ GeV,
this mode dies away quickly due to the opening of the
large $WW^*$ mode. This is the primary reason for the
limitation to a low Higgs mass ($m_h<140$) 
at a muon collider \cite{sher,hfactory}.

\begin{figure}[tb]
\mbox{\psfig{file=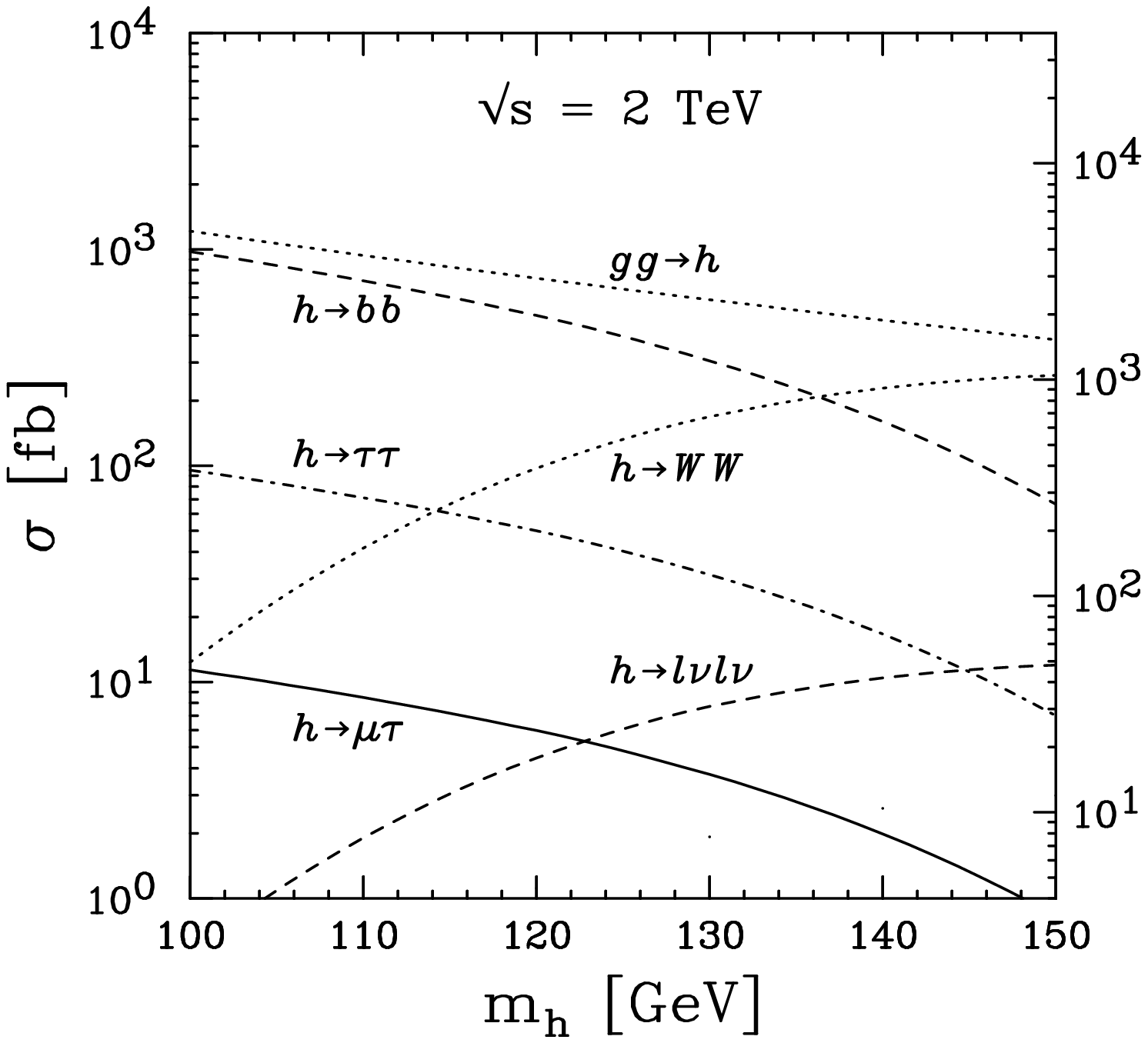,angle=90,width=4.3cm,height=5cm}
\psfig{file=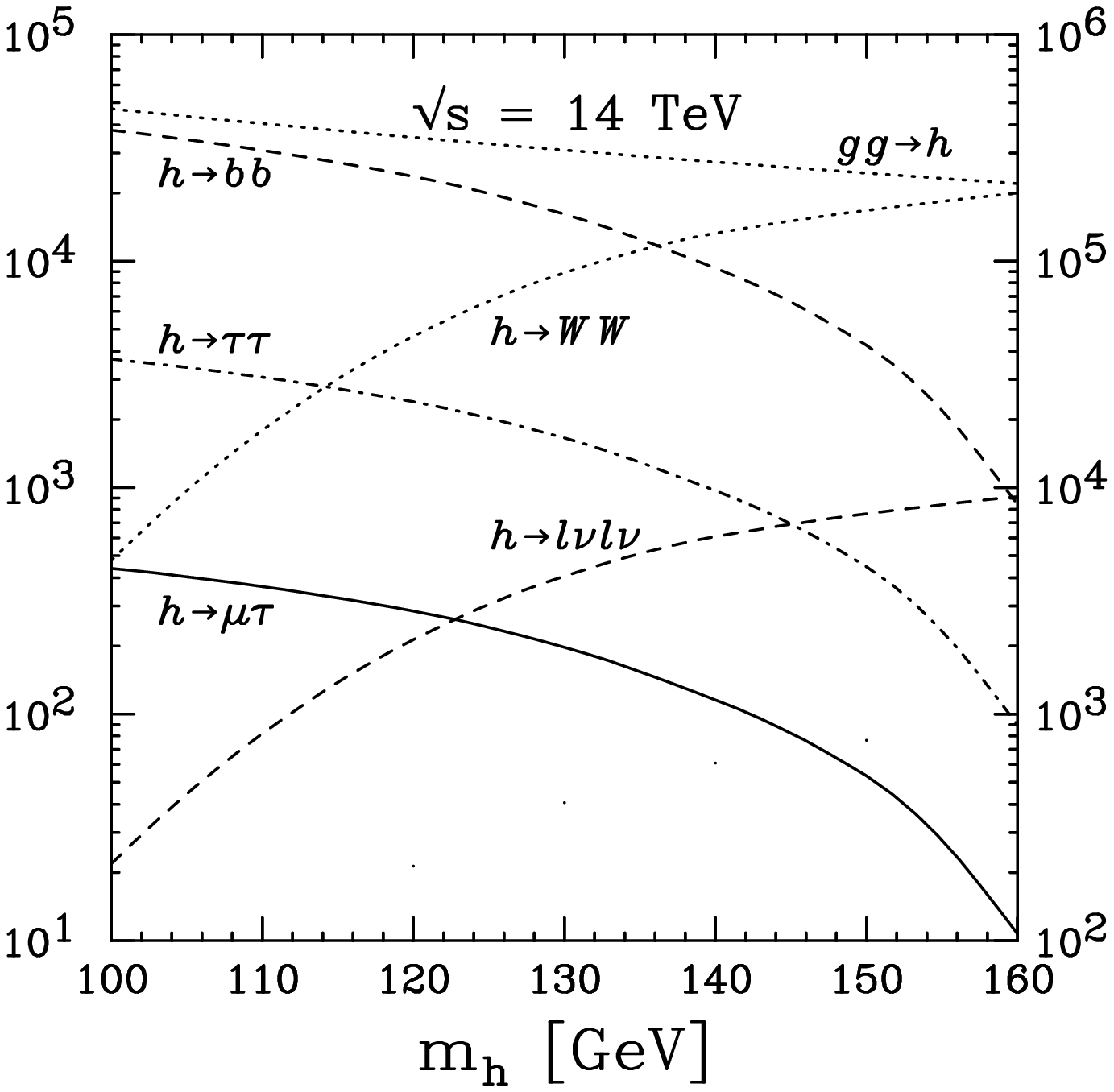,angle=90,width=4.0cm,height=5cm}}
\caption{The Higgs boson production cross-section via
 gluon-fusion versus $m_h$ at the (a) Tevatron 
and (b) LHC.
The solid curve is for the $\mt$ mode, assuming
$\kmt=1$. The scales on the right-hand side give the number of 
events expected for (a) 4 $\fbi$ 
at the Tevatron and (b) 10 $\fbi$ at
the LHC. Various subsequent decay modes
$\tau^+\tau^-$, $\ww$ and $\ww\to \ll$ are 
depicted for comparison. }
\label{SIG}
\end{figure}

In Fig.~\ref{SIG} we show the total cross section 
for $gg\to h$ as well as the final states from the
$h$ decay versus $m_h$ at the (a) Tevatron and (b) LHC. 
The production is SM-like as we take $\kappa_{tt}=1$.
We normalize our signal cross section to include
next-to-leading order QCD corrections \cite{spira},
and use the CTEQ4M distribution functions \cite{cteq4m}.
The scales on the right-hand side give the number of events 
expected for 4 $\fbi$ at the Tevatron (the 2 $\fbi$ luminosity 
at the CDF and D0 detectors are combined)
and 10 $\fbi$ at the LHC. 
We see that for the $\mh$ range of $110-140$ GeV and $\kmt=1$,
there may be about $10-40$ events produced at the Tevatron 
and $100-4000$ events at the LHC.

\vskip 0.1in
\noindent
\underline{$h\to \mu\tau$ signal and SM backgrounds}

The signal final state $\mu\tau$ is quite unique: two
flavor-changing charged leptons back-to-back in the
transverse plane without much hadronic activity.
To estimate the observability of the signal
in hadron collider environments, we consider the $\tau$
to decay to an electron or (at least one charged) hadrons,
excluding the mode to a muon. We do not require 
explicit $\tau$ tagging in the analysis. We simulate the detector
coverage at the Tevatron (LHC) by imposing some ``basic cuts''
\begin{eqnarray}
p_T^\mu>20\ {\gev},\ p_T^\pm >10\ {\gev},\ |\eta|<2\ (2.5),
\label{Basic}
\end{eqnarray}
where $p_T^\mu\ (p_T^\pm)$ is the transverse momentum
for the muon (charged track and other observable hadrons
from $\tau$ decay), and $\eta$ is their pseudo-rapidity. 
We further simulate the detector energy resolutions
at the Tevatron \cite{runii}
\begin{eqnarray}
\Delta E_j/E_j &=& 0.8/\sqrt E_j \quad {\rm for\ hadrons},\nonumber\\ 
\Delta E_e/E_e &=& 0.2/\sqrt E_e \quad {\rm for\ electrons},
\end{eqnarray}
and at the LHC \cite{LHC}
\begin{eqnarray}
\Delta E_j/E_j &=& 0.65/\sqrt E_j \oplus 0.05\quad {\rm for\ 
hadrons},\nonumber\\ 
\Delta E_e/E_e &=& 0.1/\sqrt E_e 
\oplus 0.005\quad {\rm for\ electrons}.
\end{eqnarray}
The muon is required to be well isolated and
we neglect the $p_T^\mu$ smearing.
We finally veto extra jets in the range
\begin{equation}
p_T^j > 20\ {\gev},\ |\eta^j|<3
\end{equation}
to maximally preserve the signal kinematics.

Although the lepton flavor-changing signal is quite
spectacular, it is not background-free. The leading
SM backgrounds include the Drell-Yan (DY) process
\begin{eqnarray}
\label{DY}
pp(\bar p) &\to& Z(\gamma^*) \to \tau^+\tau^-\to \mu\nu_\mu\nu_\tau\ \tau, 
\end{eqnarray}
and $W^+W^-$ pair production ($WW$)
\begin{eqnarray}
\label{WW}
pp(\bar p) &\to& W^+W^- \to \mu\nu_\mu\ \tau \nu_\tau. 
\end{eqnarray}
The background processes are calculated with the full 
SM matrix elements at tree level including spin correlations
of gauge boson decays. QCD corrections as $K$-factors for the total
production rates are also taken into account \cite{kDY}.
With the basic cuts of Eq.~(\ref{Basic}), 
the backgrounds turn out to be very large. 
The results are given by the entries under ``basic cuts''
in Tables~\ref{tabI} and
\ref{tabII} for the Tevatron and LHC, respectively.

There are several distinctive kinematical features for the 
signal that we can exploit to discriminate it from the 
backgrounds. First, the missing neutrinos from $\tau$ decay
are collimated along the charged track since the $\tau$'s
are ultra-relativistic. Thus, for the signal, the missing
transverse momentum ($p_T^{miss}$) is along the charged
track direction and is essentially back-to-back with 
respect to the muon  $\phi(\mu,\pm)\approx 180^\circ$. 
This is not the case for the $WW$ background. Secondly,
the muons in the signal are stiff $p_T^\mu\sim m_h/2$
as a result of the two-body Higgs decay; while the secondary
tracks and hadrons from $\tau$ decay are softer. If we
define momentum imbalance
\begin{equation}
\Delta p_T=p_T^\mu-p_T^\pm,
\label{Dpt}
\end{equation}
we expect that it would be positive for the signal
if the momentum measurements were perfect. This
variable turns out to be very powerful in separating
the DY background. We now define the ``refined cuts'' as
\begin{equation}
\phi(\mu,\pm) > 160^\circ,\ 
\Delta p_T>0,\ 
p_T^\mu>m_h/5.
\label{refine}
\end{equation}

The most important aspect for the signal observation is
reconstruction of the Higgs boson mass. This is quite feasible
for the signal under consideration. This can be done with
the following steps: (1) define the missing transverse momentum
$p_T^{miss}$ as the imbalance from the observable particles
(which is $\Delta p_T$ in Eq.~(\ref{Dpt}) for the signal case);
(2) reconstruct the $\tau$ transverse momentum ${{\vec p}_T}^\tau= 
{{\vec p}_T}^\pm + {{\vec p}_T}^{miss}$, and the longitudinal
component $p_z^\tau=p_z^\pm(1 + p_T^{miss}/p_T^\pm)$; (3) form the $\mt$
invariant mass $m_{\mt}^2=(p^\mu+p^\tau)^2$. This mass
variable should be sharply peaked at $m_h$ for the signal,
broadly peaked around $M_Z$ for the DY background, and rather
smooth over a large range for the $WW$ background. Indeed,
with the proper energy smearing, we find the reconstructed 
Higgs mass peak within a 5 GeV range.
The results are summarized in Tables~{\ref{tabI}} and \ref{tabII}
for the Tevatron and LHC, respectively.
The entries under ``refined cuts'' give the cross sections 
including the cuts of Eq.~(\ref{refine}). 
The signal-to-background ratio $S/B$ within a 5 GeV window
for $m_{\mt}$ is shown next. The last rows illustrate the statistical
significance $S/\sqrt{B}$ for the Tevatron with 20 $\fbi$
(CDF and D0 combined) and for the LHC with 10 $\fbi$.

\begin{table}[t]
\begin{tabular}{l|ccccc}
$\sigma$ [fb] & \multicolumn{5}{c}{$\mh$ [GeV]} \\
\hline
    & 100& 110& 120& 130& 140\\
\hline
basic cuts    &&&&&                       \\
\hline
signal & 6.5& 5.0& 3.6& 2.3& 1.3\\
DY        &  \multicolumn{5}{c}{$1.4\times 10^4$} \\
WW           &  \multicolumn{5}{c}{380} \\
\hline
refined cuts  &&&&&                               \\
\hline
signal & 5.5& 4.2& 3.0& 1.9& 1.0\\
DY [pb] & $7.6$& $6.6$& $5.6$& $4.7$ & $3.8$\\  
WW  & 60& 59& 58& 57& 55\\    
\hline
$S/B$& ${5.4\over 25}$& ${4.1\over 14}$& ${2.9\over 9.0}$& 
${1.9\over 6.4}$& ${1.0\over 4.9}$\\
$S/\sqrt{B}$ (20 fb$^{-1}$) & 4.9& 4.9& 4.5& 3.4& 2.0
\end{tabular}
\caption{Signal $h\to \mu\tau$ and SM background cross sections
at the 2 TeV Tevatron for $\mh=100-140$ GeV and $\kmt=1$ 
after different stages of 
kinematical cuts. The signal statistical significance $S/\sqrt{B}$ is
presented for 20 fb$^{-1}$.}
\label{tabI}
\end{table}

\begin{table}[t]
\begin{tabular}{l|ccccccc}
$\sigma$ [fb] & \multicolumn{7}{c}{$\mh$ [GeV]}\\
\hline
& 100& 110& 120& 130& 140& 150& 160\\
\hline
basic cuts    &&&&&&&                  \\
\hline
signal & 230& 200& 160& 120& 69& 32& 6.6\\
DY      &  \multicolumn{7}{c}{$8.9\times 10^4$}\\
WW            &  \multicolumn{7}{c}{4000}\\
\hline
refined cuts  &&&&&&&                                 \\
\hline
signal & 200& 170& 130& 94& 56& 26& 5.3\\
DY [pb]  & $48$& $42$& $36$& $30$& $24$& $19$& $14$\\
WW  & 700& 700& 690& 680& 670& 650& 630\\    
\hline
$S/B$& ${190 \over 160}$& ${160 \over 91}$& ${130 \over 63}$& ${91 \over 47}$& 
${54 \over 37}$& ${25 \over 30}$& ${5.1 \over 25}$\\    
$S/\sqrt{B}$ (10 fb$^{-1}$) & 47& 54& 52& 42& 28& 15& 3.2
\end{tabular}
\caption{Signal $h\to \mu\tau$ and SM background cross sections
at the 14 TeV LHC for $\mh=100-160$ GeV and $\kmt=1$ 
after different stages of 
kinematical cuts. The signal statistical significance $S/\sqrt{B}$ is
presented for 10 fb$^{-1}$.}
\label{tabII}
\end{table}

\vskip 0.1in
\noindent
\underline{Discussion and conclusion}

\begin{figure}[t]
\mbox{\psfig{file=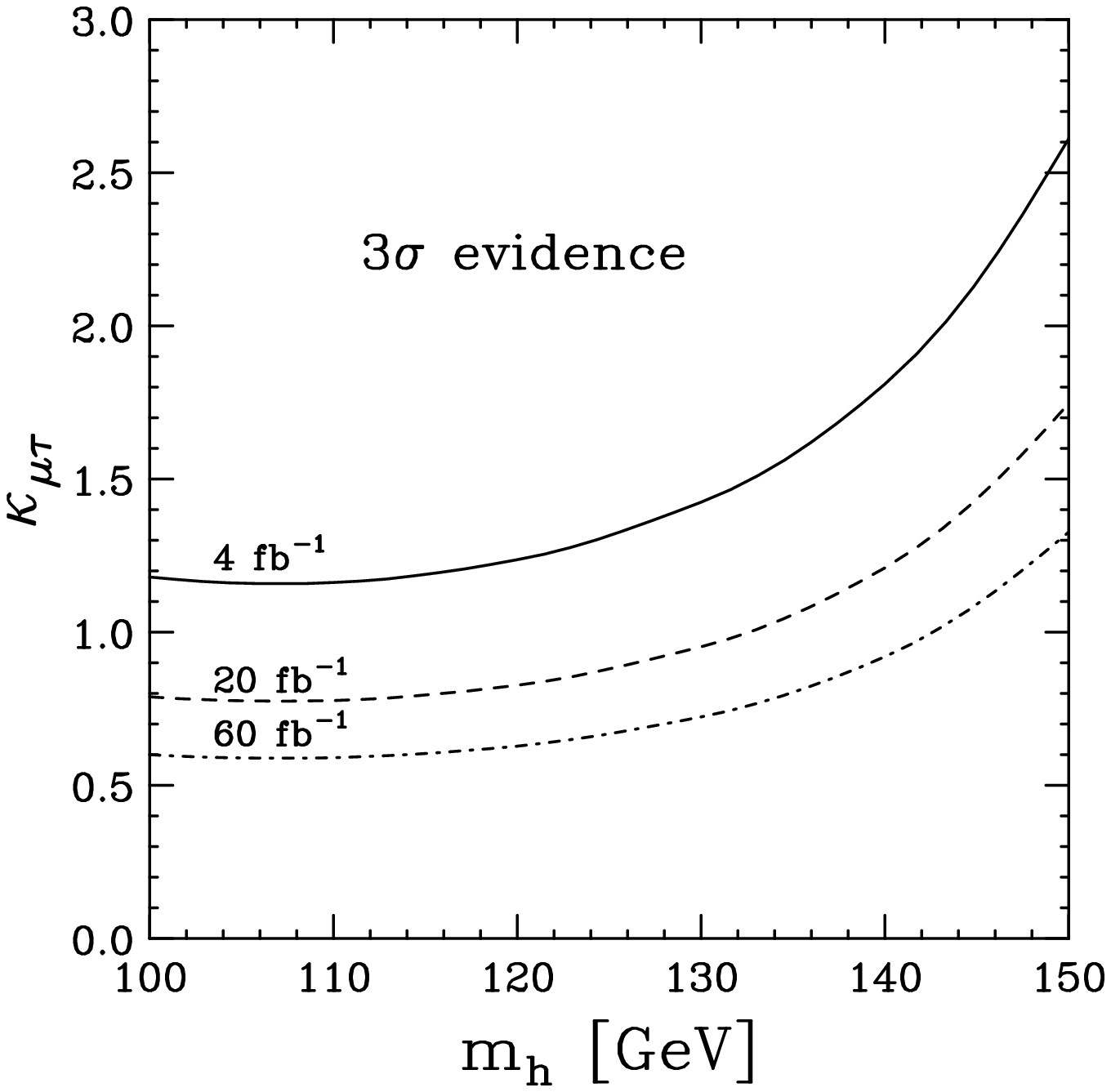,angle=90,width=4.24cm,height=5cm}
\psfig{file=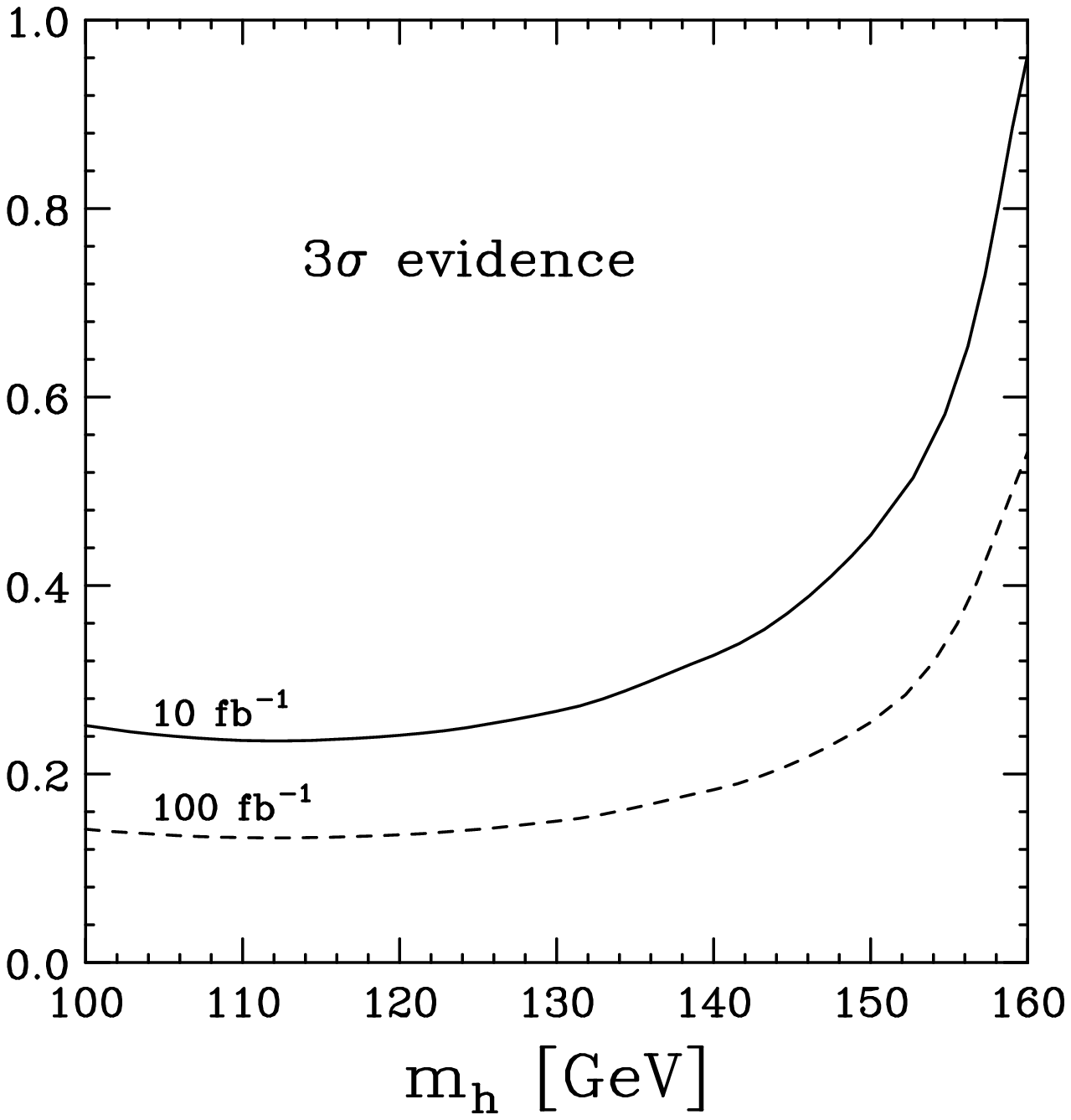,angle=90,width=4.05cm,height=5cm}}
\caption{The value of $\kmt$ at which the signal yields
a 3$\sigma$ statistical evidence, 
versus $m_h$ at the (a) Tevatron and (b) LHC
for several luminosities.
}
\label{kreach}
\end{figure}

\begin{figure}[t]
\mbox{\psfig{file=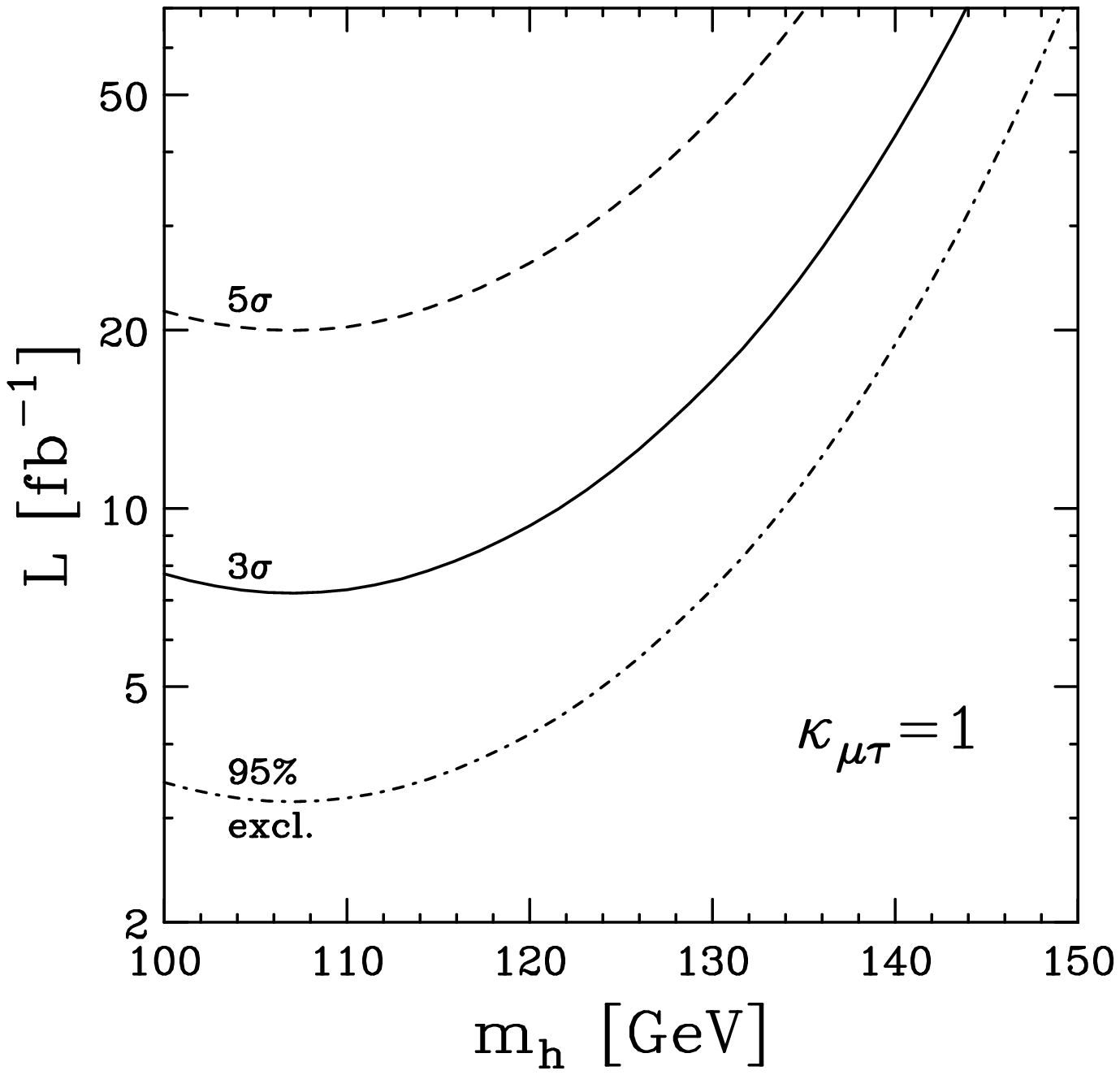,angle=90,width=4.2cm,height=5cm}
\psfig{file=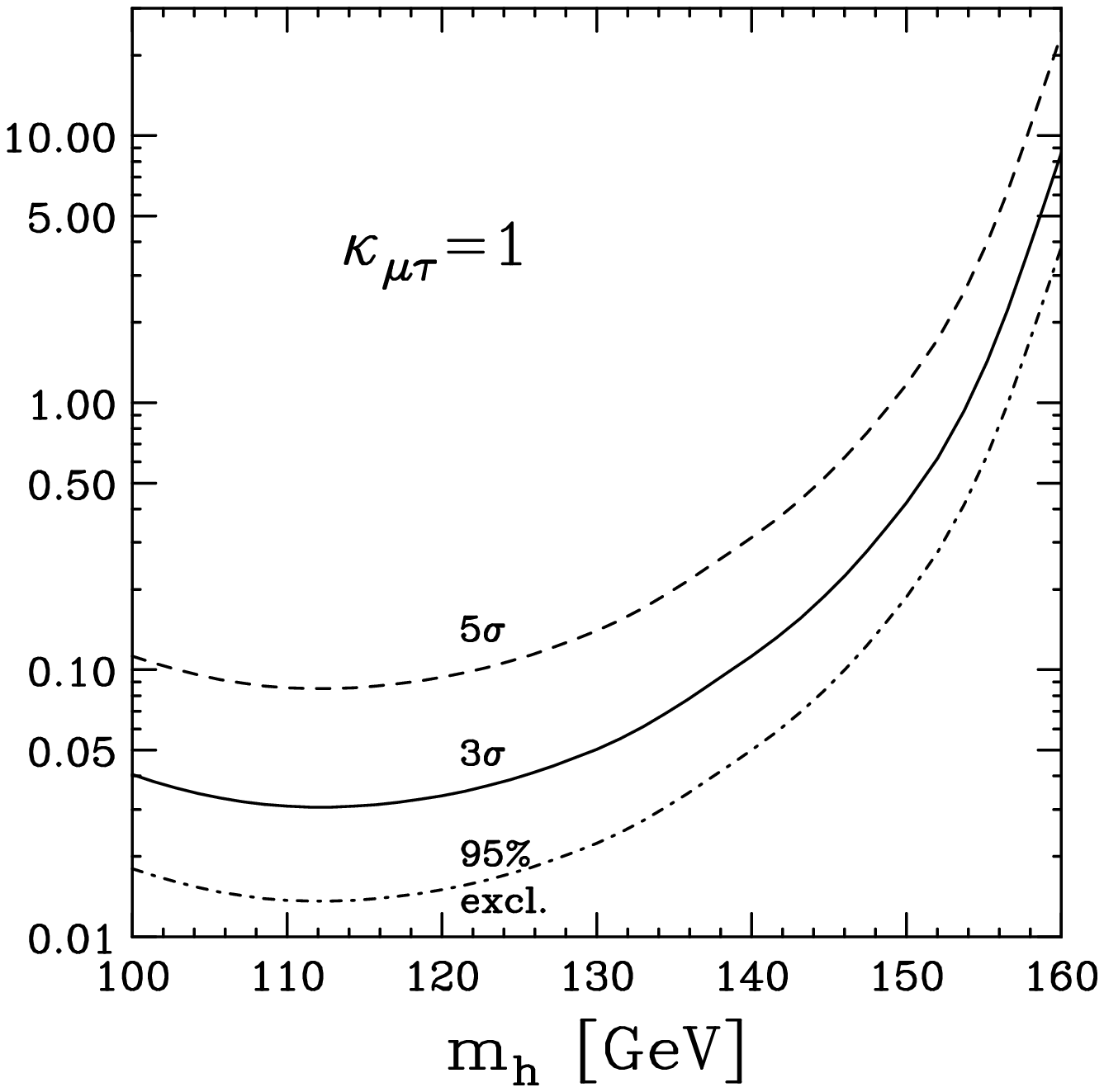,angle=90,width=4.1cm,height=5cm}}
\caption{Integrated luminosity needed to reach a 
2$\sigma\ (95\%$ exclusion),
3$\sigma$  and 5$\sigma$ signal 
versus $m_h$ at the (a) Tevatron and (b) LHC
for $\kmt=1$.
}
\label{lreach}
\end{figure}

So far, for our signal discussion, we have chosen the
coupling parameter as $\kmt=1$ for illustration. 
From a model-building point of view, it is natural
for $\kmt$ to be of order unity, while the upper
bound from low energy constraint is about 10.
Generically, the cross section scales like $\kmt^2$.
We explored to what value of this coupling the
signal would yield a 3$\sigma$ evidence statistically 
near the Higgs mass peak. Figure \ref{kreach} 
shows $\kmt$ versus $m_h$ at the (a) Tevatron 
and (b) LHC for several luminosities. We see
that at Run IIa where a luminosity of 4 $\fbi$ 
is expected combining CDF and D0 data, 
$\kmt\sim 1.2-1.8$ can be reached for 
$m_h\lsim 140$ GeV. With a higher luminosity
of 30 $\fbi$ per detector, one can reach a coupling
of $0.6-0.9$. At the LHC, the sensitivity is
significantly improved and a signal for $\kmt\sim 0.15$
would even be observable with 100 $\fbi$. Assuming
$\kmt\approx 1$, the reach could go beyond 
$m_h\approx 160$ 
GeV, in contrast to the accessible limit
$m_h\lsim 140$ 
GeV at a muon collider \cite{sher}. 
Similarly, one can ask how much luminosity is needed
to reach a certain level of observation. Note
that the statistical significance scales
like $S/\sqrt B \sim \kmt^2 \sqrt L$. The
results are summarized in Fig.~\ref{lreach}, where
a 2$\sigma\ (95\%$ confidence level exclusion), 
3$\sigma$ and 5$\sigma$ signals are
illustrated versus $m_h$ at the (a) Tevatron and (b) 
LHC for $\kmt=1$.
Due to the large number of signal events near 
the $m_h$ peak at the LHC (see Table~\ref{tabII}),
the statistical accuracy of determining a coupling
$\kmt\sim {\cal O}(1)$ can be at a few percent level
with only $L=10\ \fbi$. 
Note that strictly speaking, 
all the bounds quoted here apply to the product $\kappa_{tt} \kmt$.
We have implicitly assumed $\kappa_{tt}=1$ throughout.

In summary,
we have studied the observability for a lepton flavor-changing 
decay of a Higgs boson $h\to \mu\tau$ at the upgraded Tevatron and 
the LHC. At the
Tevatron, the unique signature may serve as the Higgs 
discovery channel, yielding a 3$\sigma$ signal for
$m_h\sim 110$ GeV and $\kmt\sim 1.2$ with 4 $\fbi$
(CDF and D0 combined), surpassing expectations for Higgs boson
searches in the SM and in a large parameter region of the MSSM. 
The sensitivity will be 
greatly improved at the LHC, probing as small a coupling
as $\kmt\sim 0.15$ or determining $\kmt\sim {\cal O}(1)$ 
better than a few percent accuracy, and extending the reach to
$m_h\sim 160$ GeV, beyond the coverage at
a muon collider.

{\it Acknowledgments}:
This work was supported in part by a DOE grant
No. DE-FG02-95ER40896 and in part by the Wisconsin Alumni 
Research Foundation.

\end{document}